\documentclass[a4paper,12pt]{amsart}

\usepackage[left=1in, right=1in, top = 1.5in, bottom = 1.5in]{geometry}
\usepackage[utf8]{inputenc}
\usepackage[all]{xy}
\usepackage{amssymb}
\usepackage{amsmath}
\usepackage{csquotes}
\usepackage{subfig}
\usepackage{enumerate}
\usepackage{graphicx}
\usepackage{enumerate}
\usepackage{amsaddr}
\usepackage{color}
\usepackage{mwe}
\usepackage{subfig}
\usepackage{appendix}

\usepackage{soul}
\usepackage{xcolor}

\usepackage{tikz}
\usetikzlibrary{arrows.meta,shapes}
\usetikzlibrary{arrows,shapes.arrows,shapes.geometric,shapes.multipart,
decorations.pathmorphing,positioning,shapes.swigs,}

\usepackage{setspace}
\theoremstyle{definition}
\newtheorem{example}{Example} 

\MakeOuterQuote{"}\EnableQuotes
\DeclareUnicodeCharacter{00A0}{ }

\title[]{Translating questions to estimands in randomized clinical trials with intercurrent events}

\begin{document}
\author{Mats J. Stensrud$^{1}$, Oliver Dukes$^2$} \address{ $^1$ Department of Mathematics, École Polytechnique Fédérale de Lausanne, Switzerland \\
$^2$ Department of Applied Mathematics, Statistics and Computer Science, Ghent University, Belgium \\
}
\maketitle

\begin{abstract}
    Intercurrent (post-treatment) events occur frequently in randomized trials, and investigators often express interest in treatment effects that suitably take  account of  these events. A naive conditioning on intercurrent events does not have a straight-forward causal interpretation, and the practical relevance of other commonly used approaches is debated. In this work, we discuss how to formulate and choose an estimand, beyond the marginal intention to treat effect, from the point of view of a decision maker and drug developer. In particular, we argue that careful articulation of a practically useful research question should either reflect decision making at this point in time or future drug development. Indeed, a substantially interesting estimand is simply a formalization of the (plain English) description of a research question. A common feature of estimands that are practically useful is that they correspond to possibly hypothetical but well-defined interventions in identifiable (sub)populations. To illustrate our points, we consider five examples that were recently used to motivate consideration of principal stratum estimands in clinical trials. In all of these examples, we propose alternative causal estimands, such as conditional effects, sequential regime effects and separable effects, that correspond to explicit research questions of substantial interest. Certain questions require stronger assumptions for identification. However, we  highlight that our proposed estimands require less stringent assumptions than estimands commonly targeted in these settings, including principal stratum effects. 
\end{abstract}

\section{Introduction}

Defining, interpreting and identifying causal effects in the presence of an intercurrent (post-treatment) event is not straightforward, even in a randomized controlled trial (RCT). It is well-known that randomization of a baseline treatment does not ensure identification of estimands that are (implicitly or explicitly) defined conditional on a post-treatment variable. Moreover, a naive contrast of outcomes conditional on the post-treatment variable does not have a clear causal interpretation. 


In this manuscript, we discuss key issues concerning the formulation and choice of a causal estimand in settings with intercurrent events. Our first message is that an explicit research question should always precede the choice of estimand: it is the question that motivates the estimand, not the estimand that motivates the question. While this point may seem obvious and has been pointed out several times in the causal inference literature, numerous studies fail to give such a motivation. As a result, the interpretation -- and practical relevance -- of these analyses is ambiguous. 

To illustrate the importance of carefully defining a causal estimand, we revisit five recent examples \cite{bornkamp2020principal} inspired by the ICH addendum  \cite{ich2019addendum}. These examples concern settings where the intercurrent event is non-absorbing and non-interveneable. As will become clear in Sections \ref{sec: estimands description}-\ref{sec: identification conditions}, such events do not prevent a classical intention to treat (ITT) analysis from being valid and identified. However, decisions makers and drug developers will often be interested in additional questions that translate to different estimands. We argue that the causal estimands of interest in these secondary analyses should be selected on a case-by-case basis. 

Furthermore, we clarify that the translation of a research question into a formal estimand 
is separate from the question of its identification, i.e. whether and how it can be expressed as a functional of distributions of observables.
Establishing plausible conditions for identification can be considered a distinct -- but important -- task, which should be carefully conducted \textit{after} choosing the causal estimand. The final (distinct) task is estimation, for which one may choose from more or less parametric or robust approaches. In other words, we find it helpful to distinguish between the following three tasks of data analyses: (i) translation of the research question into a formal causal estimand, (ii) assessing conditions for identification of the causal estimand 
and (iii) estimation of the causal estimand from observed data. In this article, we will focus on task (i), but also briefly consider tasks (ii) and (iii). 

We have structured our arguments as follows. In Section \ref{sec: intro examples}, we briefly introduce the intercurrent event settings in the five clinical examples from Bornkamp et al \cite{bornkamp2020principal}. In Section \ref{sec: estimands description}, we describe estimands that have been suggested for causal inference in clinical settings with intercurrent events.\footnote{Section \ref{sec: estimands description} is a stand-alone section that can be passed over by readers who are already familiar with these estimands.} In Section \ref{sec: all examples}, we revisit the five clinical examples and map the subject matter questions  -- as described in plain English by the investigators -- to causal estimands. In Section \ref{sec: identification conditions}, we review conditions that allow us to identify these estimands in a RCT, followed by a brief description on how to estimate them in Section \ref{sec: estimation}. In Section \ref{sec: discussion}, we give a short discussion.    


\section{Clinical examples}
\label{sec: intro examples}

\example[Multiple sclerosis]
\label{ex intro: multiple sclerosis}
Multiple sclerosis (MS) is a progressive neurological disease. Initially most patients have a phase with relapses followed by recoveries, and eventually the patients transform to a secondary phase with less frequent relapses and gradual decline. There is major interest in developing new treatments that delay or prevent disease progression. For example, the EXPAND study was a randomised clinical trial that assigned the drug siponimod versus placebo to patients in the secondary phase, where the primary estimand was the onset of confirmed disability progression \cite{kappos2018siponimod}. Siponimod was shown to delay the onset of disability progression compared to placebo, and it was also shown to reduce the frequency of relapses. These primary results raised the question "whether a treatment effect (on the onset of disability progression) would be present in patients that would not experience relapses" \cite{bornkamp2020principal, magnusson2019bayesian}. Here, experiencing relapses is an intercurrent event. 

\example[Treatment effects in early responders]
\label{ex intro: early resp}
A biomarker is a variable that quantifies a biological state. Treatment effects on biomarkers, such as high sensitivity c-reactive protein \cite{ridker2017antiinflammatory},  can serve as early predictors of treatment effects on clinical outcomes. The ASA/EFSPI oncology estimand working group \cite{bornkamp2020principal} further suggested that "biomarkers or early readouts can be useful to investigate whether an investigational
medicine works as intended on a biological level". Thus, a motivation for studying biomarker responses in RCTs seems to be elucidation of causal mechanisms (on a biological level), which can indicate whether the drug acts as intended through particular causal pathways. Here, reaching a biomarker threshold is an intercurrent event. 

\example[Antidrug antibodies (ADA) for targeted oncology trials]
\label{ex intro: ada}
Immunotherapies are increasingly used to treat several cancers. Yet, there is concern that some immunotherapies can trigger the production of antidrug antibodies (ADAs), which in turn can reduce the treatment effect of the immunotherapies on, say, overall survival \cite{enrico2020antidrug}. Indeed, "ADAs may be directed against immunogenic parts of the drug and may affect its efficacy or safety, or they may bind to
regions of the protein which do not affect safety or efficacy, with little to no clinical effect" \cite{bornkamp2020principal}. Here, the production of ADAs is an intercurrent event. 

\example[Impact of exposure on overall survival]
\label{ex intro: survival}
In drug trials, individuals who are given the same dose of a drug can still have different concentration of the active substance of this drug in the blood (serum). The concentration of the active substance can be important for the treatment effect. A RCT that assigned Trastuzumab for gastric cancer showed that those patients in the quartile with the lowest drug concentration had worse overall survival (OS) compared to the other quartiles \cite{bang2010trastuzumab}. Researchers at the Food and Drug Administration \cite{yang2013combination}, subsequently raised the question "whether the lower OS is due to low drug concentration
or to disease burden" (see also \cite{cosson2014population, bornkamp2020principal}). Here, the drug concentration is considered to be an intercurrent event \cite{bornkamp2020principal}. 

\example[Prostate cancer prevention]
\label{ex intro: prostate cancer}
There is major interest in developing drugs that prevent development of diseases, such as cancer. In particular, finasteride was shown to reduce the rates of prostate cancer in an RCT \cite{thompson2003influence}. While there were lower rates of cancer in the finasteride arm, those who developed cancer in the finasteride arm had, on average, more aggressive cancers than those assigned to placebo. This conditional association, however, does not necessarily have a causal interpretation --- it could just be that the more aggressive cancers cannot be prevented by finasteride while mild ones can be prevented. Thus, after establishing that finasteride reduces the rate of prostate cancer, Bornkamp et al \cite{bornkamp2020principal} suggested a secondary analysis to assess whether "the effect of finasteride on the severity of prostate cancer among those men who would be diagnosed with prostate cancer regardless of their treatment assignment".  Here, being diagnosed with prostate cancer is an intercurrent event. 

\subsection{Intercurrent events that are non-absorbing and non-interveneable.}
\label{sec: non-abs non-int}
In Examples \ref{ex intro: multiple sclerosis}-\ref{ex intro: prostate cancer}, the intercurrent event does not render the outcome of interest ill-defined. We can evaluate the effect of siponimod vs.\ placebo on disability progression in the entire study population in Example \ref{ex intro: multiple sclerosis}, whether or not patients have relapses. Similarly, we can evaluate the effects of canakinumab on MACE (Example 2), immunotherapies on cancer progression (Example 3), Trastuzumab on gastric cancer (Example 4) and finasteride on prostate cancer incidence (Example 5), whether or not individuals experience the intercurrent events. Thus, we can study total effect estimands like \ref{eq: ate} in all of the examples, but nevertheless there may be interest in endpoints that take into account the intercurrent events. 

Furthermore, we do not know about any intervention that fixes the intermediate event in Examples \ref{ex intro: multiple sclerosis}-\ref{ex intro: prostate cancer}. In particular, we do not have any current or future treatment that fixes MS patients to relapse or not relapse. 

Thus, Examples \ref{ex intro: multiple sclerosis}-\ref{ex intro: prostate cancer} are characterized by intercurrent events that are \textit{non-absorbing} and \textit{non-interveneable}. Other intercurrent events do not necessarily share these features. For example, Michels et al \cite{michiels2020novel} studied treatment effects in clinical studies where patients sometimes received rescue medications in response to worsening of a disease. Here, the use of rescue medication was considered to be an intercurrent event, and it is easy to conceive an intervention to give or reject rescue treatment.\footnote{Despite the fact that these interventions are easy to conceptualize, they may be impossible to implement in practice, e.g.\ due to ethical reasons.} Furthermore, some outcomes are often considered to be ill-defined in the presence of an intercurrent event. For example, many researchers consider quality of life to be ill-defined after death, where death can be considered to be an intercurrent event. Whether or not events are absorbing and interveneable has implications for the plausibility of estimands, as we discuss in more detail in Section \ref{sec: estimands description}.

\section{Notation and Estimands for settings with intercurrent events}
\label{sec: estimands description}


\subsection{Average total effects}
\label{sec: sequential trial estimands}
The primary 
estimand in most randomized trials is the (average) total effect of treatment assignment.
This is called the \textit{treatment policy strategy} estimand in the ICH addendum \cite{ich2019addendum}[page 8], and it coincides with the intention-to-treat (ITT) effect in case of non-adherence. For simplicity, we will consider average effects, often on the additive scale, in the remainder of this manuscript. Such averages of individual level effects are often the most relevant for decision making and have a clear causal interpretation. However, our conceptual points are also valid for other counterfactual contrasts. 

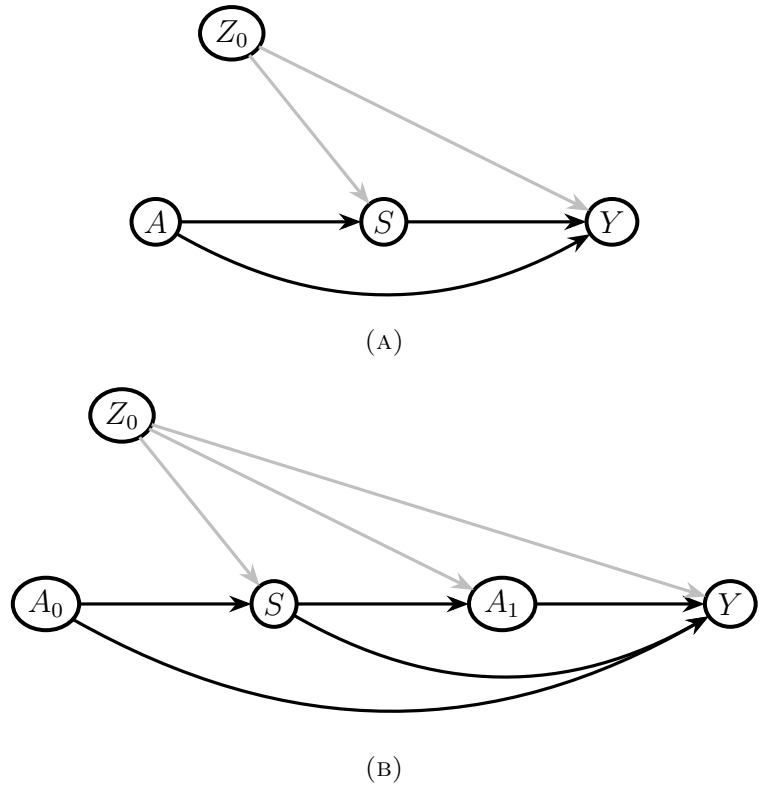
\begin{figure}
    \centering
    \subfloat[]{
\begin{tikzpicture}
\tikzset{line width=1.5pt, outer sep=0pt,
ell/.style={draw,fill=white, inner sep=2pt,
line width=1.5pt},
swig vsplit={gap=5pt,
inner line width right=0.5pt}};
\node[name=A,ell,  shape=ellipse] at (0,0) {$A$}  ;
\node[name=S,ell,  shape=ellipse] at (3,0) {$S$}  ;
    \node[name=Y,ell,  shape=ellipse] (Y) at (6,0) {$Y$};
    \node[name=Z0,ell,  shape=ellipse] at (1,2.5) {$Z_0$};
\begin{scope}[>={Stealth[black]},
              every edge/.style={draw=black,very thick}]
	\path [->] (A) edge (S);
	\path [->] (A) edge[bend right] (Y);
    \path [->] (S) edge (Y);
    \path [->,>={Stealth[lightgray]}] (Z0) edge[lightgray]  (S);
    \path [->,>={Stealth[lightgray]}] (Z0) edge[lightgray]  (Y);
\end{scope}
\end{tikzpicture}
} \\
    \subfloat[]{
\begin{tikzpicture}
\tikzset{line width=1.5pt, outer sep=0pt,
ell/.style={draw,fill=white, inner sep=2pt,
line width=1.5pt},
swig vsplit={gap=5pt,
inner line width right=0.5pt}};
\node[name=A0,ell,  shape=ellipse] at (0,0) {$A_0$}  ;
\node[name=A1,ell,  shape=ellipse] at (6,0) {$A_1$}  ;
\node[name=S,ell,  shape=ellipse] at (3,0) {$S$}  ;
    \node[name=Y,ell,  shape=ellipse] (Y) at (9,0) {$Y$};
    \node[name=Z0,ell,  shape=ellipse] at (1,2.5) {$Z_0$};
\begin{scope}[>={Stealth[black]},
              every edge/.style={draw=black,very thick}]
	\path [->] (A0) edge (S);
	\path [->] (A0) edge[bend right] (Y);
	\path [->] (A1) edge (Y);
	\path [->] (S) edge (A1);
    \path [->] (S) edge[bend right] (Y);
    \path [->,>={Stealth[lightgray]}] (Z0) edge[lightgray]  (S);
    \path [->,>={Stealth[lightgray]}] (Z0) edge[lightgray]  (Y);
    \path [->,>={Stealth[lightgray]}] (Z0) edge[lightgray]  (A1);
\end{scope}
\end{tikzpicture}
}
\caption{Causal DAGs that describe a randomized trial where baseline treatment $A$ is randomly assigned (A), and a sequential randomized trials where $A_0$ and $A_1$ are randomly assigned, where the assigment of $A_1$ depends on the intercurrent event $S$.}
    \label{fig: dag}
\end{figure}

To formally define this estimand, consider a study where  individuals are randomly assigned to a binary treatment $A\in \{0,1\}$ at baseline; while, strictly, it is the assignment that is randomized, we will simply speak of the randomized treatment $A$ in the following. Let $Y$ be the outcome of interest measured at a fixed time $t>0$, and let $S$ be an indicator of an intercurrent event that occurs at a time $b$ such that $b < t$.  A corresponding causal DAG is show in Figure \ref{fig: dag}A, where randomization of $A$ ensures there are no common causes of $A$ and $Y$. Let superscripts denote potential outcomes, such that $Y^{a}$ and $S^{a}$ denote the potential outcome of interest and the post-treatment event, respectively, had an individual, possibly contrary to the fact, been assigned $A=a$, where $a \in \{0,1\}$. The average total effect of the 
treatment $A$ on the outcome $Y$ is a contrast  
\begin{align}
& \mathbb{E} ( Y^{a=1}) \text{ vs. }  \mathbb{E} ( Y^{a=0}).
\label{eq: ate}
\end{align}

The average total effect \eqref{eq: ate} compares the average outcome in the population had everyone been treated ($a=1$) versus not treated ($a=0$); it ignores any intercurrent events. In all of the examples of  Section \ref{sec: intro examples}, the average total effect of treatment assignment is well-defined and meaningful. It describes the effect of assigning siponimod vs. placebo on disability (Example 1), canakinumab vs. placebo on MACE (Example 2), immunotherapies vs. placebo on cancer progression (Example 3), Trastuzumab vs. placebo on gastric cancer in patients receiving chemotherapy (Example 4) and finasteride vs. placebo on prostate cancer incidence (Example 5), without considering any intercurrent events. The intercurrent events in these examples are such that the treatment / control and the outcome remain meaningful with or without the intercurrent event. In other  situations this total effect will not necessarily be well-defined, as discussed in the ICH addendum, such as "discontinuation of assigned treatment, use of an additional or alternative treatment, drop-out and terminal events such as death" \cite{ich2019addendum}[Section A.1] (See also \cite{stensrud2021discussion, young2021identified}). Non-adherence, or changes in treatment such as rescue treatment, may imply a substantial modification of intended treatment which is why an ITT analysis is sometimes regarded as unsatisfactory;  while competing risks or drop-out may make the outcome impossible to occur or be measured, in which case the effect measure is not well-defined. 


Consider now a setting where treatments are \textit{sequentially} given at multiple times, which motivates the study of sequential treatment effects \cite{robins1986new, hernan2018causal}. For example, let $A$ be an immunotherapy, and suppose that the doctor every week (sequentially) recommends whether a patient should initiate, continue or discontinue the therapy. To fix ideas, suppose that treatment can be given at baseline (time $0$) and one subsequent point in time (time $1$), and let $A_0$ and $A_1$ be treatment indicators at these times (Figure \ref{fig: dag}B). The causal effect comparing a sequential treatment strategy that fixes $A_0=a_0$ and then $A_1=a_1$ versus an alternative strategy that fixes $A_0=a_0'$ and then $A_1=a_1'$ is 
\begin{align}
    \mathbb{E}(Y^{a_0,a_1})  \text{ vs. }  \mathbb{E}(Y^{a_0',a_1'}),
    \label{eq: seq ate}
\end{align}
where $a_0,a_1,a_0',a_1' \in \{0,1\}$. Importantly, the sequential trial estimand is often of interest, even in clinical trials that assign a single treatment strategy at baseline.  For example, formal (causal) definitions of per protocol estimands \cite{hernan2018cautions}, which arguably are useful in a range of practical settings such as pragmatic trials \cite{hernan2018causal, hernan2017per}, often require specification of sequential treatment regimes: the per protocol estimands evaluate the effect of taking treatment, as described in the protocol, at \textit{every} point in time versus not taking treatment at \textit{every} point in time. Alternatively, we might consider the effect of treatment received at baseline, if no one (or everyone) received treatment subsequently. The \textit{hypothetical strategy estimand} referred to in the ICH addendum \cite{ich2019addendum} can thus also be viewed as a sequential trial estimand.

Estimand \eqref{eq: seq ate} is a special case of a broader class of so-called dynamic sequential treatment regimes. These regimes are called dynamic, because the sequential treatment decisions can depend on each patients previous treatment and other characteristics. In other words, the treatment decision at time $k$ can be a function of (time-varying) covariates up until time $k$, and these covariates can include the intercurrent event status and treatment that have been received previously. For example, a decision of continuing a medical treatment at a time $k$ may depend on the clinical history up until time $k$, previous treatments that were received and potential side-effects. 

Let $g \in \mathcal{G}$ be a regime that fixes the treatment at two points in time: First $A_0$ is set to $a_0$ and subsequently $A_1$ is set to $a_1 = f_g(a_0, s) $ where $f_g (\cdot) $ is a deterministic function of the first treatment $a_0$ and the intercurrent event $s \in \{0,1\}$: that is, a regime $g$ where the treatment at time $1$ depends on whether the patient received the treatment at time $0$ and the status of the intercurrent event $S$ before time $1$. For example, suppose $g$ is defined such that only individuals who received treatment at time $0$ and did not experience the intercurrent event will receive treatment at time $1$.

We can define a contrast of two regimes $g, g' \in \mathcal{G}$, 
\begin{align}
& \mathbb{E} ( Y^{g}) \text{ vs. }  \mathbb{E} ( Y^{g'}).
\label{eq: dymanic treat}
\end{align}

The total effect \eqref{eq: ate} and the sequential trial estimands \eqref{eq: seq ate}-\eqref{eq: dymanic treat} do not quantify the mechanisms by which the treatment affects the outcome $Y$. In particular, these estimands neither quantify effects conditional on nor mediated through the intercurrent events. On the other hand, these estimands are immediately relevant for designing practically feasible treatment regimes for the existing treatment $A$. Clinicians and patients will be particularly interested in the optimal regime $g^* \in \mathcal{G}$ that leads to the most favourable expected clinical outcome, i.e.\ 
\begin{align}
    & g^*  \equiv \underset{g \in \mathcal{G}}{\arg \max } \  \mathbb{E}(Y^g) \label{eq: optimal g}. 
\end{align}

The average total effects \eqref{eq: ate}  and the sequential trial estimands \eqref{eq: seq ate}-\eqref{eq: dymanic treat} compare (counterfactual) outcomes under different treatment assignments in the entire study population. These estimands can easily be modified to average effects conditional on (functions of) observed pre-treatment variables $Z$, such as 
$$
\mathbb{E} ( Y^{g} \mid Z) \text{ vs. }  \mathbb{E} ( Y^{g'} \mid Z), 
$$
and we discuss such conditional effects in more detail in Section \ref{sec: all examples}. Alternatively, we can define conditional causal effects of the second treatment, given the first treatment and other events before time $k$.  The causal effect of assigning $A_1=a_1$ versus $A_1=a_1'$ among those who received baseline treatment $a_0$ and experienced the intercurrent event $s$ is
\begin{align}
    \mathbb{E}(Y^{a_0,a_1}   \mid S^{a_0}=s).  \text{ vs. }  \mathbb{E}( Y^{a_0,a_1'}  \mid S^{a_0}=s).
    \label{eq: seq trial}
\end{align}
The estimand \eqref{eq: seq trial} only quantifies effects of the second, but not the first, treatment assignment. Furthermore, this estimand is restricted to the subpopulation that experienced the intercurrent event under treatment $a_0$.


\subsection{Principal Stratum Effects}
\label{sec: principal stratum effects}
The term principal stratification was coined by Frangakis and Rubin as follows \cite{frangakis2002principal}: "\textit{Principal stratification with respect to a posttreatment
variable is a cross-classification of subjects defined by the joint potential values of that posttreatment variable
under each of the treatments being compared}." This idea was introduced by Robins \cite{robins1986new},\footnote{Unlike Frangakis and Rubin \cite{frangakis2002principal}, Robins \cite{robins1986new} expressed a sceptical view of the practical importance of these estimands.} when he considered counterfactual outcomes in individuals who would survive, regardless of treatment assignment.\footnote{The principal stratum effects in this survival setting are often denoted survivor average causal effects \cite{tchetgen2014identification,egleston2009estimation}.} 
To fix ideas about principal stratum effects in a clinical trial setting, consider a randomized trial where a binary treatment $A\in \{0,1\}$ is assigned at baseline. Let $Y \equiv Y(t)$ be the outcome of interest measured at a fixed time $t>0$, and let $S \equiv S(b)$ be an indicator of a post-treatment event defined at a fixed time $b$, where $0 < b \leq t$.\footnote{Thus, for the ease of exposition, we have simplified $Y$ and $S$ to be time-fixed variables, similar to Bornkamp et al \cite{bornkamp2020principal}. However, these variables are time-varying in many practical settings.} 

The additive principal stratum effect in the stratum defined by $S^{a=1}=s, =S^{a=0}=s'$ is
\begin{align}
& \mathbb{E} ( Y^{a=1} - Y^{a=0} \mid 
S^{a=1}=s, =S^{a=0}=s').
\label{eq: princ strat}
\end{align}
Note that there are \textit{joint potential outcomes} in the conditioning set of \eqref{eq: princ strat}; this effect is defined in the subset of individuals characterized by the counterfactual intercurrent events  $S^{a=1}=s$ and $S^{a=0}=s'$. In the example of a vaccine trial (suggested in the ICH addendum \cite{ich2019addendum}), one might consider the treatment effect in those who would be infected regardless of vaccine assignment (say, $S^{a=1}=s, =S^{a=0}=s$). Because this subpopulation has the same intercurrent event regardless of treatment, the principal effect can be interpreted as both a direct effect of the vaccine (outside of infection) and a total effect.

The fact that \eqref{eq: princ strat} is defined with respect to joint potential outcomes has raised concern in the statistical causal inference literature \cite{robins1986new, robins2010alternative,stensrud2020conditional, robins2007discussions,joffe2011principal, dawid2012imagine, vanderweele2011principal, pearl2011principal}. In general, these joint potential outcomes cannot be observed in the same individual, and therefore the principal stratum effects are defined in an unobserveable subpopulation that may not even exist, although it may sometimes be possible to obtain informative bounds on the size of the principle strata. 

The ICH addendum \cite{ich2019addendum} uses a broader definition of principal stratum effects, which encompasses estimands that condition on combinations (unions) of principal strata. In particular, the contrast 
\begin{align}
& \mathbb{E} ( Y^{a=1} - Y^{a=0} \mid 
S^{a=1}=s),
\label{eq: princ strat ich}
\end{align}
where the conditioning set is the union $ (S^{a=1}=s, =S^{a=0}=1) \cup (S^{a=1}=s, =S^{a=0}=0)  $ is, in addition to \eqref{eq: princ strat}, included in the ICH definition of principal stratum estimands. The conditioning set in \eqref{eq: princ strat ich}, that is, the union of principal strata, can be identified from a randomized trial. For example, in a vaccine trial we can look at the subpopulation in the vaccine arm who were infected. Yet, the interpretation of \eqref{eq: princ strat ich} is not straightforward, because it does not necessarily express a direct effect.\footnote{Unless additional assumptions are imposed.} In particular, \eqref{eq: princ strat ich} possibly quantifies effects both through and outside of the intercurrent event $S$ unless it is assumed that the treatment $A$ does not affect $S$ \cite{stensrud2020conditional}.

\subsection{Causal effects that reflect mechanisms}
\label{sec: mechanism estimands}
In the presence of intercurrent events, investigators often raise questions about the mechanisms by which the treatment affects the outcome of interest; that is, whether the treatment affects the outcome of interest through or outside of the intercurrent events. To clarify why such questions are of substantial interest, our experience is that the investigators give stories about modified treatments that leverage certain mechanisms of the original treatment \cite{robins2010alternative, robins2020interventionist, didelez2018defining, stensrud2019separable, stensrud2020generalized, stensrud2021discussion, young2021identified}. The motivation seems to be that by understanding the causal mechanisms by which the current treatment affects the outcome of interest, we can motivate new, improved treatments in the future. Here we will define a class of mechanistic estimands called separable effects \cite{robins2010alternative, didelez2018defining, stensrud2019separable, stensrud2020generalized, stensrud2020conditional, robins2020interventionist}, inspired by the seminal treatment decomposition idea from Robins and Richardson \cite{robins2010alternative}, that are particularly helpful in this setting. To fix ideas about separable effects, we start with an example. 

\example[Statins]
\label{ex intro: statins}
Statins are one of the most commonly prescribed drug classes in the world. They are successful because they reduce the risk of cardiovascular disease in a broad range of individuals with various clinical histories.  The main mechanism by which statins reduces cardiovascular risk is lowering of low-density lipoprotein (LDL) cholesterol. However, a substantial fraction of patients who take statins experience muscle symptoms, which represent a big hurdle and can lead to treatment discontinuation \cite{buettner2008prevalence}. More recently, new classes of drugs have been developed to specifically lower LDL levels like statins, but nevertheless differ from statins in the effects through other biological pathways. In particular, protein convertase subtilisin/kexin type 9 (PCSK9) inhibitors selectively reduce LDL levels, but PCSK9 inhibitors do not exert effects through other biological pathways that is affected by statins (and can lead to muscle symptoms). Indeed, PCSK9 inhibitors have been successfully shown to reduce cholesterol levels in patients with muscle related statin intolerance \cite{nissen2016efficacy}. 


\subsubsection{Separable effects are effects of modified treatments}
The separable effects are designed to target effects of (future) treatments, which selectively exert effects through certain (desirable) causal pathways similarly to the original treatment (e.g.\ the LDL lowering induced by statins), but also selectively avoid other (undesirable) causal pathways (e.g.\  non-selective pathways that can lead to muscle pain). In Section \ref{sec: all examples} we consider the potential role separable effects in the examples from Section \ref{sec: intro examples}, where new (improved) drugs remain to be discovered. 

More abstractly, the separable effects are defined with respect to modified treatment components, $A_Y$ and $A_S$, which are linked to the original treatment $A$ in the following way: when $A_Y$ and $A_S$ are set to the same value $a$, the effects of giving this combination of modified treatments ($A_Y = A_S = a$), is the same as the effects of giving the original treatment $A=a$ for $a=0,1$ (see Appendix \ref{app: more formally on separable effects} for a more formal derivation and \cite{robins2010alternative, didelez2018defining, stensrud2019separable, robins2020interventionist} for more details). This modified treatment assumption holds when the original treatment $A$ can be decomposed into two components $A_Y$ and $A_S$. However, this assumption can also hold even if $A$ cannot be physically decomposed \cite{stensrud2020conditional}. When the components $A_Y$ and $A_S$ are given individually, they exclusively target certain causal pathways (see Appendix \ref{app: more formally on separable effects} for details). 

In practice, a study of separable effects should be motivated by a scientific story about modified treatments $A_Y$ and $A_S$; the investigators should clarify why modified treatments are of scientific interest. One reason could be to develop improved treatments in the future. For example, let $A$ be statin therapy and $Y$ be cardiovascular risk. The combination  $A_Y=1$ and $A_S=0$ can be conceived as a modified therapy (such as PCSK9 inhibitors) that, like statins, have a cholesterol lowering component $A_Y = 1$, but lacks effects on the intercurrent event such as muscle symptoms, $A_S  = 0$. 

More explicitly, we define the separable effect of the $A_Y$ component as
\begin{equation}
\Pr (Y^{a_Y=1,a_S}=1)\text{ vs. }\Pr
(Y^{a_Y=0,a_S}=1), \quad a_S \in \{0,1\},
\label{eq: A_Y sep}
\end{equation}
which quantifies the causal effect of the $A_Y$ component on the risk of the outcome of under an intervention that assigns $A_S=a_S$.  Similarly,
\begin{equation}
\Pr (Y^{a_Y,a_S=1}=1)\text{ vs. }\Pr
(Y^{a_Y,a_S=0}=1), \quad a_Y \in \{0,1\},
\label{eq: A_S sep}
\end{equation}%
quantifies the causal effect of the $A_S$ component on the event of interest under an intervention that assigns $A_Y=a_Y$. The separable effect \eqref{eq: A_Y sep} quantifies the direct effect of the treatment on the outcome of interest, but this direct effect is different from the principal stratum estimand because it is defined in the entire study population. Similarly, \eqref{eq: A_S sep} quantifies the indirect effect of the treatment on the outcome of interest through the intercurrent event. This decomposition makes separable effects  natural for studying mechanisms, and, unlike the sequential trial or principal stratification estimands, it offers a coherent notion of an indirect effect.



\subsubsection{Conditional separable effect}
In settings with intercurrent events, researchers often express interest in estimands in subgroups defined by the intercurrent event. This is particularly relevant when the intercurrent event is absorbing, such that an outcome  may only be well defined conditional on the event status (e.g. when there is censoring due to death). Following Stensrud et al \cite{stensrud2020conditional}, under the assumption that $A_Y$ partial isolation \eqref{def: Ay partial iso} holds, we can also define a conditional separable effect as
\begin{align}
&\mathbb{E} ( Y^{a_Y=1,a_S}  - Y^{a_Y=0,a_S} \mid
S^{a_S}=s ).
\label{eq: conditional sep eff 1}
\end{align}
The conditional separable effect is the average causal effect of the (modified) treatment $A_Y$ on $Y$ when all individuals are assigned the other (modified) treatment $A_S=a_S$ in those individuals who do not experience the intercurrent event under $A_S=a_S$ (regardless of their value of $A_Y$). 
 
\subsection{Relations between estimands}
Like conventional mediation estimands \cite{robins2010alternative, robins2020interventionist}, such as natural direct and indirect effects \cite{robins1992identifiability}, the (marginal) separable effects quantify the mechanisms by which the treatment affects the outcome of interest \cite{robins2010alternative, robins2020interventionist, stensrud2019separable, didelez2018defining, stensrud2020generalized}. Unlike the conventional mediation estimands, the separable effects do not require specification of any intervention on the intermediate event. It is not clear that such interventions on the intercurrent events exist in any of Examples 1-5. Furthermore, a feature of the separable effects is that they are defined with respect to modified treatments, and therefore they can be directly relevant in a drug development setting where the interest is tailoring new treatments to include/exclude certain pathways. However, as we return to in Section \ref{sec: identification conditions}, conventional mediation estimands and separable effects are often identified by the same functionals of the observed data. The implication of this is that the existing computer software for conventional mediation estimands can often be used to calculate separable effects. 

Moreover, the conditional separable effects are related to principal stratum effects \cite{stensrud2020conditional}. Indeed, a conditional separable effect is restricted to subjects who have a certain value of the intercurrent event under the modified treatment $a_S$. Assuming that $A_Y$ does not exert effects on $S$,  \eqref{eq: conditional sep eff 1} is equal to $\mathbb{E} ( Y^{a_Y=1,a_S} - Y^{a_Y=0,a_S} \mid S^{a}=s)$ which targets the same subgroup considered in the PS estimand \eqref{eq: princ strat} rather than \eqref{eq: princ strat ich}. However, it also explicitly quantifies a treatment effect that acts \textit{outside} of the intercurrent event (a \textit{direct} effect on the event of interest not mediated by the intercurrent event) in this subgroup. Furthermore, under an additional monotonicity assumption that $S^{a=1}_{ } \leq S^{a=0}_{ }$, which also must be justified on scientific grounds (and is often implausible),  Stensrud et al \cite{stensrud2020conditional} showed that a conditional separable effect is in fact identical to a principal stratum effect defined with respect to joint outcomes \eqref{eq: princ strat}.\footnote{Without the monotonicity assumption, the separable effects may still map to combined principal strata estimands like \eqref{eq: princ strat}.}

\section{Translating Research Questions to Estimands}
\label{sec: all examples}

We now revisit the five examples discussed in Section \ref{sec: intro examples}. We map research questions to their corresponding estimands. We find that all of these estimands are interventionists estimands that, at least in principle, could be studied in a (future) randomized trial \cite{richardson2013single, robins2020interventionist}.

\setcounter{example}{0}
\example[Multiple sclerosis (cont.)]
\label{sec: MS}
In a secondary analysis of the EXPAND trial \cite{kappos2018siponimod}, Magnusson et al \cite{magnusson2019bayesian} studied a principal stratum estimand like \eqref{eq: princ strat}, but defined on the risk ratio scale,
\begin{align}
&\frac{  \mathbb{E} (Y^{a=1}\mid 
S^{a=1}=S^{a=0}=1) }{ \mathbb{E} (Y^{a=0}\mid 
S^{a=1}=S^{a=0}=1)}  ,
\label{eq: princ strat spec}
\end{align}
where $Y$ indicates confirmed disability progression (at some time $t$) and the conditioning set $S^{a=1}=S^{a=0}=1$ indicates having no relapses \textit{regardless} of treatment assignment at baseline. Later this estimand was also advocated by the ASA/EFSPI oncology estimand working group \cite{bornkamp2020principal}. Estimand \eqref{eq: princ strat spec} quantifies  disability under siponimod versus no treatment in subject who would not relapse
under both siponoind and no treatment.\footnote{We consider relapse at a time $s < t$ before assessment of the disability outcome $Y \equiv Y(t)$.} 
 
One motivation for studying principal stratum effects, like \eqref{eq: princ strat spec}, of siponimod was  "\textit{understanding the effect of siponimod on progression occurring independently of relapses}" \cite{magnusson2019bayesian}.   
The relevance of effects in those who would not experience relapse regardless of treatment assignment is nevertheless unclear; as we cannot observe this subset of the population of unknown size, it cannot be a direct target population for a new drug in the future. 

On the contrary, treatment effects \textit{outside} of relapse are of interest if the investigators consider the opportunity of leveraging or avoiding these particular effects in future, refined drugs. For example, based on a biochemical evidence, the drug developers may have reasons to believe that the current drug exerts effects through different (biological) pathways, which have differential effects on relapse and disability progression. Questions about such mechanisms of action motivate the consideration of separable effects \cite{robins2010alternative, didelez2018defining, stensrud2019separable, stensrud2020generalized, robins2020interventionist},\footnote{We use the term separable effect broadly to denote interventionists estimands for causal mechanisms \cite{robins2010alternative, robins2020interventionist, didelez2018defining, stensrud2019separable, stensrud2020generalized, stensrud2020conditional}, which cover classical mediation settings, competing events and truncation by death.} which explicitly targets the mechanism by which a treatment exerts effects on an outcome \cite{robins2010alternative, didelez2018defining, stensrud2019separable, stensrud2020generalized, stensrud2020conditional,young2021identified, stensrud2021discussion}. In particular, suppose that siponimod exerts effects on disability $(Y=1)$ through a pathway avoiding relapse $(S=1)$, and suppose that we could create a new drug $(A_Y=1)$ that exclusively targets this pathway, but does not act via the pathways by which siponimod reduces relapse in MS patients. Analogously, we could, in principle, imagine a different treatment $(A_S=1)$ that exclusively exerts effects on relapse, similar to siponimod, but does not exert effects on disability outside of the relapse pathway. This immediately motivates separable effects estimands, defined with respect to a hypothetical trial where we assign the modified treatments $A_Y$ and $A_S$,
\begin{align}
&\mathbb{E} ( Y^{a_Y=1,a_S}) \text{ vs. }  \mathbb{E} (Y^{a_Y=0,a_S}),
\label{eq: marg sep eff 1}
\end{align}
which is defined (marginally) in the full study population. The contrast \eqref{eq: conditional sep eff 1} quantifies the effect of siponimod versus the new drug that exclusively exerts effects on relapse. Thus, if the contrast \eqref{eq: conditional sep eff 1} is equal to zero, then there is no effect of siponimod outside of its effect on relapse. If the contrast is different from zero, then siponimod also exerts effects outside of relapse on disability. 

Magnusson et al \cite{magnusson2019bayesian} also state that there was particular interest in the treatment effect \textit{``among the subgroup of patients for whom relapses would be absent during the study''}. We may therefore consider a conditional separable effects estimand like \eqref{eq: conditional sep eff 1},
\begin{align*}
&\mathbb{E} ( Y^{a_Y=1,a_S} \mid
S^{a_S}=0 ) \text{ vs. }  \mathbb{E} (Y^{a_Y=0,a_S} \mid
S^{a_S}=0).
\end{align*}
This conditional separable effect quantifies the effect of siponimod versus the new drug that exclusively exerts effects on relapse, but, unlike \eqref{eq: marg sep eff 1}, it is confined to those who would not relapse on siponimod. 

So far we have discussed estimands that quantify effects through certain causal mechanisms, which e.g.\ can motivate the development of future drugs. However, if the aim is to support labelling decisions of siponimod itself, as also suggested in \cite{magnusson2019bayesian}, other estimands seems more relevant. In particular, doctors and regulators are often interested in whether the treatment effect varies across subgroups of patients. Such subgroup effects can only be useful for practical decision making if the subgroups are  observed \textit{before} the decision is made, which is not the case for principal stratum effects. On the other hand, simple average treatment effects conditional on a set of measured baseline covariates $Z$,
\begin{align}
    \mathbb{E}(Y^{a=1} - Y^{a=0} \mid Z=z),
    \label{eq: simple baseline cond estimand}
\end{align}
is of immediate interest. Whereas the conditional average treatment effect is easy to define (and identify), summarizing subgroup effects across covariates $Z=z$ is not trivial. One way to summarize (coarsen) conditional effects is to define an auxiliary variable based on expected outcomes under treatment. In particular, instead of studying effects in principal strata, we could study effects for groups of patients who are \textit{likely to be} in a principal stratum of interest \cite{joffe2007defining}. Following Joffe et al \cite{joffe2007defining}, define the principal score $\mu^a (Z) =  P(S=a \mid Z) $ \cite{hill2002differential}, which in our example denotes the probability of not having relapses  under treatment $A=a$ given baseline covariates $Z$. The pair of principal scores $(\mu^{a=0} (Z) , \mu^{a=1} (Z) )$ is, unlike the principal stratum, a baseline covariate, because it is just a function of $Z$. We can now define causal effects among individuals with the same principal scores under treatment $a$, such as
\begin{align}
& \mathbb{E}(Y^{a=1} - Y^{a=0} \mid \mu^{a} (Z) = q), \text{ or } 
\label{eq: principal score}  \\
 & \mathbb{E}(Y^{a=1} - Y^{a=0} \mid \mu^{a} (Z) > q ),
\label{eq: principal score prob}
\end{align}
which are the effect of siponimod on disability among patients with probability $q$ or probability larger than $q$, respectively, of developing recurrences under treatment $A=a$. Similarly, we could study the joint principal scores, such as
 \begin{align}
 \mathbb{E}(Y^{a=1} - Y^{a=0} \mid \mu^{a=1} (Z) = q_1, \mu^{a=0}(Z) = q_0  ),
\label{eq: principal score joint}    
\end{align}
which is the effect of siponimod on disability among patients with probability $q_1$ of developing recurrences when assigned to sifonimod and  probability $q_0$ when assigned to placebo.

Unlike the principal stratum estimands, these principal score effects are identified at baseline and can thus be used to guide decision in practice. For example, a decision maker can be interested in giving a different treatment to patients with a high risk of relapses compared to patients with low risk of relapses. However, because the principal scores are just functions of $Z$, decision rules that use $\mu^{a=0}(Z)$ and/or $\mu^{a=1}(Z)$ as input will not be better than decision rules that only use $Z$ as input.\footnote{Here, "better" alludes to optimal expected outcomes $g^*  \equiv \underset{g \in \mathcal{G}}{\arg \max } \  \mathbb{E}(Y^g) $, where $g$ is a decision rule (regime). }

\example[Treatment effects in early responders (cont.)]
\label{sec: early resp}

To motivate the study of treatment effects in early responders, Bornkamp et al \cite{bornkamp2020principal} gave a subject-matter example on the effect of canakinumab (an immunotherapy) vs.\ placebo on major adverse cardiovascular
events (MACE), where the authors write that "as the mechanism of action of canakinumab is lowering inflammation, one would suspect that patients who do not achieve the biomarker threshold also have a lower benefit in terms of the time-to-event outcome".
The question concerns whether canakinumab \textit{only} exerts effects on the outcome of interest through its effects on inflammation. Knowledge of this mechanism can motivate the development of new (refined) drugs, similarly to the motivating question in Example \ref{ex intro: multiple sclerosis}. To answer this question, a separable effect can be explicitly formulated as follows: suppose that we could give a modified drug $(A_Y=1)$ in which the part of canakinumab that exerts effects on inflammation is blocked, but otherwise this drug is identical to canakinumab. Would this new drug have a beneficial effect on MACE compared to no treatment $(A_Y=0)$? 

If this new treatment has a beneficial effect, then there is a component of canakinumab that acts outside of inflammation. Specifically, this question corresponds to a separable effect \eqref{eq: conditional sep eff 1}, defined in the subset of the population who would not have a biomarker response \textit{on} treatment (i.e., those with $S^{a_S=1}=0$). Unlike the principal stratum estimand \begin{align}
& \mathbb{E} ( Y^{a=1} \mid 
S^{a=1}=s) \text{ vs. }  \mathbb{E} (Y^{a=0} \mid 
S^{a=1}=s),
\label{eq: post cond}
\end{align}  
proposed in \cite{bornkamp2020principal}, the separable effect explicitly quantifies a direct effect of treatment.


An alternative motivation for studying treatment effects in early responders is to "support the decision on treatment modifications after treatment start" \cite{bornkamp2020principal}. This motivation does not concern drug development, but rather sequential decision making, and the substantive question motivates the study of a sequential trial estimand, as described in Section \ref{sec: sequential trial estimands}. To illustrate this point in the simplest possible setting, consider a trial in which treatment decisions can be made at two time points, $t=0$ and $t=1$, and suppose that the biomarker response is know at time $t=1$ but not at time $t=0$. Suppose further that patients can be assigned canakinumab ($A(t)=1$) or no treatment ($A(t) = 0$) at times $t \in \{0,1\}$. The authors` plain English motivation suggests the counterfactual estimand
\begin{align}
    \mathbb{E}(Y^{a_0=1,a_1=1} - Y^{a_0=1,a_1=0}  \mid S^{a_0=1}=s),
    \label{eq: seq trial 2}
\end{align}
This estimand quantifies the effect of a strategy that assigns canakinumab sequentially at baseline (time $0$) and time 1 versus a strategy that assigns no treatment at baseline and time 1. In order to make inference about the sequential target estimand, we need to observe individuals that receive different treatment strategies over follow up: if not, we cannot in general identify these effects.

\example[ADA for targeted oncology trials (cont.)]
\label{sec: ada}

There is interest in whether ADAs bind to a \textit{region} of the immunotherapeutic drug that affects its efficacy. In other words, there is interest in the mechanisms by which the immunotherapy exerts effects on the clinical outcome (say, all-cause survival at a given time $t$) in the presence of ADAs \cite{ridker2017antiinflammatory, bornkamp2020principal}. Hence, despite the fact that survival is a well-defined outcome whether or not the intercurrent event (ADA production) occurs, a study of total treatment effects on survival will not provide information on the effect of treatment in the presence of ADAs. 

Nevertheless, the story about mechanisms immediately motivates a study of a separable effect, similarly to Example \ref{ex intro: multiple sclerosis} and the first question in Example \ref{ex intro: early resp}. The authors also concretely suggest (physical) components of treatment \cite{robins2010alternative, didelez2018defining, stensrud2019separable, stensrud2020conditional,stensrud2020generalized} that exert different effects: Let $A$ indicate immunoterapy ($A=1$) or no treatment ($A=0$). The immunotherapy $(A=1)$ can be decomposed into a component to which the antibody binds ($A_S=0$, a particular part of the drug), and the remaining component to which it does not bind  (say, $A_Y=1$, the remaining part of the drug). Then, the estimand in \eqref{eq: conditional sep eff 1}, conditional on $S^{a_S=1}=S^{a=1}=s$ (those who would have ADA production when treated), would quantify the effect of the component to which the antibody binds ($A_Y$) on the clinical outcome ($Y$ , say, survival after a time $t$) among those who would get ADA under treatment (See Appendix \ref{app: ex ada} for more details on the relation to Bornkamp et al's \cite{bornkamp2020principal} principal stratum estimand). 


\example[Impact of exposure on overall survival (cont.)]
There is interest in "whether the lower OS is due to low drug concentration
or to disease burden" \cite{bang2010trastuzumab, yang2013combination, bornkamp2020principal}.  
This causal question alludes to an effect of \textit{low serum concentration of the active drug} versus no treatment on overall survival; for example, a practitioner may ask whether it is sufficient to have a low serum concentration to experience a treatment effect. Alternatively, the question may allude to the effect of \textit{low} versus \textit{high} serum concentration of the active drug on overall survival; for example, a practitioner may ask whether there is a dose-response effect. Both of these effects are specifically defined with respect to an intervention on the drug concentration itself: suppose, for example, that the variable $S=1$ indicates that an individual has a drug concentration in the lowest quartile on treatment, and let $S=0$ indicate that the drug concentration is equal to 0 (the concentration under no treatment). Then, the question posed by the investigators suggests the simple conditional effect 
\begin{align}
    \mathbb{E}(Y^{s=1} - Y^{s=0} \mid X=x),
    \label{eq: baseline cond estimand}
\end{align}
where $X$ is a set of covariates that are available before the (hypothetical) decisions is made:  here, $X$ could be the a trivial random variable if we are interested in the marginal effect of $S$, or $X$ could be other variables temporary ordered before $S$, in particular it is possible that $A \subset X$. However, we emphasise that the exposure of interest in this setting is the drug concentration ($S$) in the blood, and not the drug administered ($A$). Indeed, a study conducted by researchers at the FDA \cite{yang2013combination} aimed to study whether individuals with low serum concentration of the active drug would \textit{show survival benefit} if they had a higher serum concentration \cite{yang2013combination}[[page 166].

One practical reason for studying \eqref{eq: baseline cond estimand}, which concerns the drug concentration, is to assess whether a higher \textit{dose} of the drug should preferably be administered, in particular to certain subgroups. In practice, however, these subgroups must be known \text{before} the treatment is given, which means that these subgroups \textit{cannot} be principal strata (which are defined by intercurrent events that are unobserved at the time of treatment initiation). However, principal scores, like \eqref{eq: principal score}-\eqref{eq: principal score prob}, could be used to make individualized decision; for example, it is possible to give a higher dose to those individuals who are likely to have a low serum concentration of the drug under a standard dose. 

Indeed, after the FDA study \cite{yang2013combination}, a randomized trial has been conducted to evaluate the effect of different drug doses in certain groups of patients  \cite{shah2017heloise}, where these groups were solely defined by pre-treatment variables, like $X$ in \eqref{eq: baseline cond estimand}.



\example[Prostate cancer prevention (cont.)]
The investigators expressed interest in "the effect of finasteride on the severity of prostate cancer among those men who would be diagnosed with prostate cancer regardless of their treatment assignment" \cite{bornkamp2020principal}. This sentence alludes to a principal stratum estimand, similar to the setting in Examples \ref{ex intro: multiple sclerosis}, \ref{ex intro: early resp} and \ref{ex intro: ada}. However, if the research question is whether finasteride does not just prevent mild cancers but also  leads to more aggressive cancer via a separate mechanism, this calls for a different estimand. If finasteride exerted such harmful effects, drug developers could aim to modify finasteride (or construct an alternative treatment), such that these harmful effects were avoided while preserving the preventive effect. Again, this motivates a separable effect, corresponding to estimand \eqref{eq: conditional sep eff 1}: Let $A_S=1$ be a new drug that exerts the same effects as finasteride on cancer prevention, and let $A_Y=1$ be another (potentially harmful) substance that has the same effect as finasteride on everything else except cancer progression. The question of the decision maker seems to be whether a new drug, which exerts the effect of the component $A_S=1$, but does not exert any of the other effects of finasteride (that is, $A_Y=0$), would be better than administering finasteride (now we can think of finasteride as equivalent to a drug containing the components $A_Y=1, A_S=1$). 



\section{Identification assumptions}
\label{sec: identification conditions}

Our arguments in Sections \ref{sec: estimands description} and \ref{sec: all examples} concerned the translation of a substantial research question, articulated in plain English, to a causal estimand, articulated in counterfactual notation. The arguments for choosing an estimand did not rely on the assumptions that are required to identify this estimand from the data at hand (e.g.\ from a RCT where only $A$ is randomly assigned). However, when an estimand of interest is chosen, it is crucial to understand, and critically assess, the assumptions necessary to learn about the estimand from data (i.e.\ to evaluate whether a sufficient set of identifiability conditions hold). Furthermore, when setting up a randomised trial, these assumptions can be helpful in understanding how the study needs to be designed, or which data needs to be collected.

Identifiability conditions for all of the estimands in Sections \ref{sec: estimands description} and \ref{sec: all examples} are thoroughly described in several previous works \cite{robins1986new, pearl, hernan2020causal, richardson2013single, robins2010alternative, stensrud2020conditional}. Here we informally review some important features of these conditions, which should be justified on scientific grounds. We focus on non-parametric identifiability conditions. Sometimes (strong) parametric assumptions can be invoked to obtain alternative identification conditions; that is, identification conditions that only hold under a particular (parametric) model. For example, mixture models and Bayesian methods have often been used to identify principal stratum effects, but as noted by Bornkamp et al \cite{bornkamp2020principal}, inference is then highly sensitive to the correctness of the parametric assumptions and likelihood estimators can exhibit pathological behaviour. Indeed, the practical value of these alternative parametric conditions is limited unless the investigators have convincing arguments why we should believe in the parametric assumptions (which, to our knowledge, is hardly the case in medicine). 

\subsection{The average total effect requires the weakest assumptions}
\label{sec: id ate weak}
Importantly, only the conventional average effect estimand \eqref{eq: ate} can be identified without additional assumptions from a (perfectly executed) RCT where the baseline treatment $A$ is randomly assigned, as illustrated in the simple causal directed acyclic graph (DAG) in Figure \ref{fig: dag}A. This result is also valid for average effect that are defined conditional on observed baseline covariates, like the effect \eqref{eq: baseline cond estimand} and principal score effects like \eqref{eq: principal score}-\eqref{eq: principal score prob}. All the other estimands we consider -- including principal stratum effects, separable effects and sequential treatment effects -- require additional assumptions in this setting. 

\subsection{Sequential trial estimands}
Sequential trial estimands can be identified without additional assumptions from a (perfectly executed) trial where the treatment is randomly assigned sequentially, i.e.\ at multiple points in time. However, when the treatment is only randomly assigned at baseline (e.g. in a randomised trial with treatment switching/discontinuation), additional assumptions are required \cite{robins1986new, richardson2013single, hernan2020causal}. In particular, it is necessary to adjust for common causes of the outcome $Y$ and the sequential treatment $A_k$ at time $k$. Causal graphs allow us to visualize and evaluate the independencies that ensure identification of sequential effects \cite{pearl, richardson2013single}. They are useful for translating and discussing the key assumptions with scientific collaborators, in order to assess their validity. For example, the causal DAG in Figure \ref{fig: dag}B describes a setting where the sequential trial estimands \eqref{eq: seq trial} and \eqref{eq: seq trial 2} are identified provided that $S$ and $Z_0$ are measured. 

\subsection{Separable effects require dismissible component conditions}
The separable effects also require conditional independencies (called dismissible component conditions or conditional exchangeabilities) to be satisfied \cite{robins2010alternative, didelez2018defining, stensrud2020generalized, robins2020interventionist}. Like the sequential trial estimands, these conditions require adjustment for common causes of the outcome $Y$ and the intercurrent event $S$, which may be time-varying \cite{robins2010alternative,stensrud2020generalized, stensrud2020conditional, robins2020interventionist}. The dismissible component conditions would also be violated if certain variables are directly affected by both components $A_Y$ and $A_S$.\footnote{In this setting, a recanting witness makes it impossible to identify the separable effects \cite{shpitser2013counterfactual, stensrud2020conditional, robins2010alternative, robins2020interventionist, stensrud2019separable}. However, note that recanting witnesses that preclude natural effects from being identified do not necessarily preclude a separable effect from being identified \cite{robins2010alternative, robins2020interventionist, stensrud2020generalized}.} 

For example, if $Z_0$ is unmeasured in the DAG in Figure \ref{fig: dag}A, then the identification conditions for these estimands are violated. Variables like $Z_0$ are likely to be present in a range of practical settings. In Example \ref{ex intro: multiple sclerosis}, any factor (e.g.\ related to genetic or lifestyle) that affects both relapse and disability would be classified as a $Z_0$. Unlike conventional mediation estimands, the separable effects are \textit{single world} estimands that can be identified in a future randomized trial \cite{stensrud2020conditional, stensrud2019separable, robins2020interventionist}.\footnote{Thus, identifiability conditions for separable effects can be evaluated in Single World Intervention Graphs (SWIGs) \cite{richardson2013single, robins2020interventionist, stensrud2019separable}.}

\subsection{Principal stratum estimands require cross-world conditions}
Unlike the other estimands, identification of principal stratum effects require assumptions are empirically unverifiable, i.e.\ untestable, even in principle \cite{robins2010alternative, richardson2013single, dawid2010identifying, joffe2011principal, stensrud2020conditional}. In particular, identifying the subgroup of individuals that constitutes the principal stratum in \eqref{eq: princ strat} is impossible in most practical settings without relying on such unverifiable assumptions \cite{robins2010alternative, richardson2013single}. This limits the practical relevance of these estimands \cite{robins2010alternative, robins2007discussions, dawid2010identifying, joffe2011principal, stensrud2020conditional}. Furthermore, a principal ignorability assumption is often invoked to identify principal stratum estimands such as \eqref{eq: princ strat} and \eqref{eq: princ strat ich} \cite{jo2009use,ding2017principal}. Interpreting the principal ignorability assumption is not straight-forward: it is a cross-world independence assumption, which requires the investigator to reason about (untestable) independencies between counterfactuals under different treatment assignments. It is necessary (but not sufficient) that the investigator adjusts for all common causes of the event of interest $Y$ and the intercurrent event $S$, even in a study where $A$ is randomly assigned, for principal ignorability to hold. These common causes may be time-varying, and, unlike the other estimands, the literature on principal stratum effects in the presence of time-varying confounding is scant and relies on assumptions that may be hard to justify \cite{tchetgen2014identification, stensrud2020conditional}. On the other hand, principal score estimands, which are defined with respect to predicted probabilities of belonging to a stratum, can be identified under weaker conditions (See Section \ref{sec: id ate weak}). 


\section{A note on estimation and software implementation}\label{sec: estimation}
This articles focuses on interpretation and identification, not estimation. However, we would like to emphasize that our consideration of interpretation and identification have consequences for estimation. In particular, we can exploit the relations between the different estimands to clarify when the same estimation algorithms (and computer software) can be used to estimate different types of effects. 

For example, computer software for natural direct and indirect effects can be used to estimate separable effects in any setting where they are identified by the same functionals of observed data distributions \cite{robins2010alternative, robins2020interventionist, stensrud2019separable}. More generally, each separable effects can be estimated using software for path-specific effects; there is always a path-specific effect that is identified by the same functional as a separable effect \cite{robins2020interventionist}. See \cite{valente2020causal} for a review of available software for causal mediation analysis.

Similarly, computer software for principal stratum effects \cite{bornkamp2020principal} can be used to estimate separable effects. In particular, under the restrictive assumptions that often are imposed to study principal stratum effects, including no-time-varying confounding and monotonicity, the software described in Bornkamp et al \cite{bornkamp2020principal} can be immediately applied to estimate the conditional separable effects \cite{stensrud2020conditional}.\footnote{In Bornkamp et al's code examples \cite{bornkamp2020principal}, \newline \texttt{https://oncoestimand.github.io/princ\_strat\_drug\_dev/princ\_strat\_example.html}, the algorithm in Section 5.1 will correspond to Stensrud et al's \cite{stensrud2020conditional} outcome regression estimator and the algorithm in Section 5.2 will correspond to Stensrud et al's \cite{stensrud2020conditional} weighted estimator.} 


\section{Discussion}
\label{sec: discussion}

We have clarified that certain causal estimands -- including sequential treatment effects, separable effects and conditional causal effects -- are useful in settings where principal stratum estimands are often recommended \cite{bornkamp2020principal,ich2019addendum}. These estimands correspond to different causal questions. If the investigator is ultimately interested in finding subgroups where treatment works better, conditional causal effects are of interest, but principal stratum effects are not relevant because the principal strata are not observable when the decision is made. We illustrate this in Examples \ref{ex intro: multiple sclerosis}, \ref{ex intro: early resp} and \ref{ex intro: survival}. Furthermore, separable effects can, unlike principal stratum effects, disentangle how the treatment affects the outcome outside or through the intercurrent event, as we illustrate in Examples \ref{ex intro: multiple sclerosis}, \ref{ex intro: ada} and \ref{ex intro: prostate cancer}.

All of our suggested estimands share key features: they are defined in observable subsets of the population, they correspond to explicit causal inquiries, and they can at least in principle be empirically falsified in a future experiment \cite{richardson2013single, robins2020interventionist, stensrud2019separable, stensrud2020conditional}. These common features do not arise by coincidence. We believe that any estimand that guides practical decision making is in principle verifiable in a well-characterized population, because any practical decision can be instantiated in this population. 

Our arguments can sound pedantic to readers who find the idea of principal stratum effects intuitive; even if the principal stratum effect does not precisely reflect the question of interest, one may claim that these estimands gives us some important insight. We disagree with this argument. Causal inference is a subtle exercise, and being precise about the interpretation of estimands -- and the practical questions that motivate them -- is crucial to avoid logical flaws and erroneous decisions. Thus, we encourage investigators to be explicit about why they target a particular estimand, which itself is an exercise that can sharpen arguments and thought processes \cite{robins2010alternative, stensrud2019separable}. 

When we have data from a trial where only a baseline treatment is randomly assigned, there are limitations as to how statistics can help us to discover causal mechanisms. Causal (structural) assumptions are required for this endeavour. These assumptions are not innocuous and are not guaranteed to hold. Estimates of separable effects can motivate the development of new treatment and generate new hypotheses of causal mechanisms, but the study of separable effects do not give guarantees that these components exist and can be used in future treatment. On the other hand, principal stratum effects do not guarantee that there exist (a substantial amount) of individuals that actually belongs to the principal stratum, do not allow us to characterize these individuals, and do not quantify the effect of future therapies. 

If estimands beyond average total effects are of interest in a randomized clinical trial, which often seems to be the case, our work illustrates an important point: the investigator should strive to include a rich set of pre- and post-treatment variables. This is crucial, because identification of mechanistic estimands and sequential trial estimands would require adjustment of (possibly time-varying) common causes of the intercurrent event and the event of interest. 

Finally, different estimands can, under certain assumptions, be identified by identical functionals of observed data distributions \cite{stensrud2020conditional}, and therefore estimated in the same way. Even if these estimands are identified by the same function of data, it is important that the investigators are explicit about their causal question and corresponding estimand: the interpretation of the estimand, and identification assumptions, can still differ \cite{robins2010alternative, didelez2018defining, stensrud2020conditional}, which matters when these estimands are going to be used for practical purposes.

\subsection*{Acknowledgements}
We thank Vanessa Didelez, Bjorn Bornkamp, Kaspar Rufibach and Baldur Magnusson for generous comments and discussions that have helped us improve the manuscript.

\bibliography{references}

\begin{thebibliography}{10}

\bibitem{bornkamp2020principal}
Bj{\"o}rn Bornkamp, Kaspar Rufibach, Jianchang Lin, Yi~Liu, Devan~V Mehrotra,
  Satrajit Roychoudhury, Heinz Schmidli, Yue Shentu, and Marcel Wolbers.
\newblock Principal stratum strategy: Potential role in drug development.
\newblock {\em Pharmaceutical Statistics}, 2021.

\bibitem{ich2019addendum}
ICH.
\newblock Addendum on estimands and sensitivity analysis in clinical trials to
  the guideline on statistical principles for clinical trials e9(r1).
\newblock {\em https://database.ich.org}, 2019.

\bibitem{kappos2018siponimod}
Ludwig Kappos, Amit Bar-Or, Bruce~AC Cree, Robert~J Fox, Gavin Giovannoni, Ralf
  Gold, Patrick Vermersch, Douglas~L Arnold, Sophie Arnould, Tatiana Scherz,
  et~al.
\newblock Siponimod versus placebo in secondary progressive multiple sclerosis
  (expand): a double-blind, randomised, phase 3 study.
\newblock {\em The Lancet}, 391(10127):1263--1273, 2018.

\bibitem{magnusson2019bayesian}
Baldur~P Magnusson, Heinz Schmidli, Nicolas Rouyrre, and Daniel~O Scharfstein.
\newblock Bayesian inference for a principal stratum estimand to assess the
  treatment effect in a subgroup characterized by postrandomization event
  occurrence.
\newblock {\em Statistics in Medicine}, 38(23):4761--4771, 2019.

\bibitem{ridker2017antiinflammatory}
Paul~M Ridker, Brendan~M Everett, Tom Thuren, Jean~G MacFadyen, William~H
  Chang, Christie Ballantyne, Francisco Fonseca, Jose Nicolau, Wolfgang Koenig,
  Stefan~D Anker, et~al.
\newblock Antiinflammatory therapy with canakinumab for atherosclerotic
  disease.
\newblock {\em New England journal of medicine}, 377(12):1119--1131, 2017.

\bibitem{enrico2020antidrug}
Diego Enrico, Angelo Paci, Nathalie Chaput, Eleni Karamouza, and Benjamin
  Besse.
\newblock Antidrug antibodies against immune checkpoint blockers: impairment of
  drug efficacy or indication of immune activation?
\newblock {\em Clinical Cancer Research}, 26(4):787--792, 2020.

\bibitem{bang2010trastuzumab}
Yung-Jue Bang, Eric Van~Cutsem, Andrea Feyereislova, Hyun~C Chung, Lin Shen,
  Akira Sawaki, Florian Lordick, Atsushi Ohtsu, Yasushi Omuro, Taroh Satoh,
  et~al.
\newblock Trastuzumab in combination with chemotherapy versus chemotherapy
  alone for treatment of her2-positive advanced gastric or gastro-oesophageal
  junction cancer (toga): a phase 3, open-label, randomised controlled trial.
\newblock {\em The Lancet}, 376(9742):687--697, 2010.

\bibitem{yang2013combination}
Jun Yang, Hong Zhao, Christine Garnett, Atiqur Rahman, Jogarao~V Gobburu,
  William Pierce, Genevieve Schechter, Jeffery Summers, Patricia Keegan, Brian
  Booth, et~al.
\newblock The combination of exposure-response and case-control analyses in
  regulatory decision making.
\newblock {\em The Journal of Clinical Pharmacology}, 53(2):160--166, 2013.

\bibitem{cosson2014population}
Val{\'e}rie~F Cosson, Vivian~W Ng, Michaela Lehle, and Bert~L Lum.
\newblock Population pharmacokinetics and exposure--response analyses of
  trastuzumab in patients with advanced gastric or gastroesophageal junction
  cancer.
\newblock {\em Cancer chemotherapy and pharmacology}, 73(4):737--747, 2014.

\bibitem{thompson2003influence}
Ian~M Thompson, Phyllis~J Goodman, Catherine~M Tangen, M~Scott Lucia, Gary~J
  Miller, Leslie~G Ford, Michael~M Lieber, R~Duane Cespedes, James~N Atkins,
  Scott~M Lippman, et~al.
\newblock The influence of finasteride on the development of prostate cancer.
\newblock {\em New England journal of medicine}, 349(3):215--224, 2003.

\bibitem{michiels2020novel}
Hege Michiels, Cristina Sotto, An~Vandebosch, and Stijn Vansteelandt.
\newblock A novel estimand to adjust for rescue treatment in clinical trials.
\newblock {\em arXiv preprint arXiv:2009.12052}, 2020.

\bibitem{stensrud2021discussion}
Mats~J Stensrud, Jessica~G Young, and Torben Martinussen.
\newblock Discussion on "causal mediation of semicompeting risks" by yen-tsung
  huang.
\newblock {\em Biometrics (accepted)}, 2021.

\bibitem{young2021identified}
Jessica~G Young and Mats~J Stensrud.
\newblock Identified versus interesting causal effects in fertility trials and
  other settings with competing or truncation events.
\newblock {\em Epidemiology}, 2021.

\bibitem{robins1986new}
James~M Robins.
\newblock A new approach to causal inference in mortality studies with a
  sustained exposure period—application to control of the healthy worker
  survivor effect.
\newblock {\em Mathematical modelling}, 7(9-12):1393--1512, 1986.

\bibitem{hernan2018causal}
Miguel~A Hernan and James~M Robins.
\newblock {\em Causal inference}.
\newblock CRC Boca Raton, FL:, 2018.

\bibitem{hernan2018cautions}
Miguel~A Hern{\'a}n and Daniel Scharfstein.
\newblock Cautions as regulators move to end exclusive reliance on intention to
  treat, 2018.

\bibitem{hernan2017per}
Miguel~A Hern{\'a}n, James~M Robins, et~al.
\newblock Per-protocol analyses of pragmatic trials.
\newblock {\em N Engl J Med}, 377(14):1391--1398, 2017.

\bibitem{frangakis2002principal}
Constantine~E Frangakis and Donald~B Rubin.
\newblock Principal stratification in causal inference.
\newblock {\em Biometrics}, 58(1):21--29, 2002.

\bibitem{tchetgen2014identification}
Eric~J Tchetgen~Tchetgen.
\newblock Identification and estimation of survivor average causal effects.
\newblock {\em Statistics in medicine}, 33(21):3601--3628, 2014.

\bibitem{egleston2009estimation}
Brian~L Egleston, Daniel~O Scharfstein, and Ellen MacKenzie.
\newblock On estimation of the survivor average causal effect in observational
  studies when important confounders are missing due to death.
\newblock {\em Biometrics}, 65(2):497--504, 2009.

\bibitem{robins2010alternative}
James~M Robins and Thomas~S Richardson.
\newblock Alternative graphical causal models and the identification of direct
  effects.
\newblock {\em Causality and psychopathology: Finding the determinants of
  disorders and their cures}, pages 103--158, 2010.

\bibitem{stensrud2020conditional}
Mats~J Stensrud, James~M Robins, Aaron Sarvet, Eric J~Tchetgen Tchetgen, and
  Jessica~G Young.
\newblock Conditional separable effects.
\newblock {\em arXiv preprint arXiv:2006.15681}, 2020.

\bibitem{robins2007discussions}
James Robins, Andrea Rotnitzky, Stijn Vansteelandt, Tom~Ten Have, Yu~Xie, and
  Susan Murphy.
\newblock Discussions on principal stratification designs to estimate input
  data missing due to death.
\newblock {\em Biometrics}, 63(3):650--658, 2007.

\bibitem{joffe2011principal}
Marshall Joffe.
\newblock Principal stratification and attribution prohibition: good ideas
  taken too far.
\newblock {\em The international journal of biostatistics}, 7(1), 2011.

\bibitem{dawid2012imagine}
Philip Dawid and Vanessa Didelez.
\newblock Imagine a can opener -- the magic of principal stratum analysis.
\newblock {\em The international journal of biostatistics}, 8(1), 2012.

\bibitem{vanderweele2011principal}
Tyler~J VanderWeele.
\newblock Principal stratification—uses and limitations.
\newblock {\em The international journal of biostatistics}, 7(1), 2011.

\bibitem{pearl2011principal}
Judea Pearl.
\newblock Principal stratification—a goal or a tool?
\newblock {\em The international journal of biostatistics}, 7(1), 2011.

\bibitem{robins2020interventionist}
James~M Robins, Thomas~S Richardson, and Ilya Shpitser.
\newblock An interventionist approach to mediation analysis.
\newblock {\em arXiv preprint arXiv:2008.06019}, 2020.

\bibitem{didelez2018defining}
Vanessa Didelez.
\newblock Defining causal meditation with a longitudinal mediator and a
  survival outcome.
\newblock {\em Lifetime data analysis}, pages 1--18, 2018.

\bibitem{stensrud2019separable}
Mats~J Stensrud, Jessica~G Young, Vanessa Didelez, James~M Robins, and Miguel~A
  Hern{\'a}n.
\newblock Separable effects for causal inference in the presence of competing
  events.
\newblock {\em Journal of the American Statistical Association},
  (just-accepted):1--23, 2020.

\bibitem{stensrud2020generalized}
Mats~J Stensrud, Miguel~A Hern{\'a}n, Eric J~Tchetgen Tchetgen, James~M Robins,
  Vanessa Didelez, and Jessica~G Young.
\newblock Generalized interpretation and identification of separable effects in
  competing event settings.
\newblock {\em arXiv preprint arXiv:2004.14824}, 2020.

\bibitem{buettner2008prevalence}
Catherine Buettner, Roger~B Davis, Suzanne~G Leveille, Murray~A Mittleman, and
  Kenneth~J Mukamal.
\newblock Prevalence of musculoskeletal pain and statin use.
\newblock {\em Journal of general internal medicine}, 23(8):1182--1186, 2008.

\bibitem{nissen2016efficacy}
Steven~E Nissen, Erik Stroes, Ricardo~E Dent-Acosta, Robert~S Rosenson, Sam~J
  Lehman, Naveed Sattar, David Preiss, Eric Bruckert, Richard
  {\v{C}}e{\v{s}}ka, Norman Lepor, et~al.
\newblock Efficacy and tolerability of evolocumab vs ezetimibe in patients with
  muscle-related statin intolerance: the gauss-3 randomized clinical trial.
\newblock {\em Jama}, 315(15):1580--1590, 2016.

\bibitem{robins1992identifiability}
James~M Robins and Sander Greenland.
\newblock Identifiability and exchangeability for direct and indirect effects.
\newblock {\em Epidemiology}, pages 143--155, 1992.

\bibitem{richardson2013single}
Thomas~S Richardson and James~M Robins.
\newblock Single world intervention graphs (swigs): A unification of the
  counterfactual and graphical approaches to causality.
\newblock {\em Center for the Statistics and the Social Sciences, University of
  Washington Series. Working Paper}, 128(30):2013, 2013.

\bibitem{joffe2007defining}
Marshall~M Joffe, Dylan Small, Chi-Yuan Hsu, et~al.
\newblock Defining and estimating intervention effects for groups that will
  develop an auxiliary outcome.
\newblock {\em Statistical Science}, 22(1):74--97, 2007.

\bibitem{hill2002differential}
Jennifer Hill, Jane Waldfogel, and Jeanne Brooks-Gunn.
\newblock Differential effects of high-quality child care.
\newblock {\em Journal of Policy Analysis and Management: The Journal of the
  Association for Public Policy Analysis and Management}, 21(4):601--627, 2002.

\bibitem{shah2017heloise}
Manish~A Shah, Rui-hua Xu, Yung-Jue Bang, Paulo~M Hoff, Tianshu Liu, Luis~A
  Herr{\'a}ez-Baranda, Fan Xia, Amit Garg, Mona Shing, and Josep Tabernero.
\newblock Heloise: Phase iiib randomized multicenter study comparing
  standard-of-care and higher-dose trastuzumab regimens combined with
  chemotherapy as first-line therapy in patients with human epidermal growth
  factor receptor 2--positive metastatic gastric or gastroesophageal junction
  adenocarcinoma.
\newblock {\em Journal of Clinical Oncology}, 35(22):2558--2567, 2017.

\bibitem{pearl}
Judea Pearl.
\newblock {\em Causality: Models, Reasoning and Inference 2nd Edition}.
\newblock Cambridge University Press, 2000.

\bibitem{hernan2020causal}
Miguel~A Hern{\'a}n and James~M Robins.
\newblock Causal inference: what if.
\newblock {\em Boca Raton: Chapman \& Hill/CRC}, 2020, 2020.

\bibitem{shpitser2013counterfactual}
Ilya Shpitser.
\newblock Counterfactual graphical models for longitudinal mediation analysis
  with unobserved confounding.
\newblock {\em Cognitive science}, 37(6):1011--1035, 2013.

\bibitem{dawid2010identifying}
Philip Dawid, Vanessa Didelez, et~al.
\newblock Identifying the consequences of dynamic treatment strategies: A
  decision-theoretic overview.
\newblock {\em Statistics Surveys}, 4:184--231, 2010.

\bibitem{jo2009use}
Booil Jo and Elizabeth~A Stuart.
\newblock On the use of propensity scores in principal causal effect
  estimation.
\newblock {\em Statistics in medicine}, 28(23):2857--2875, 2009.

\bibitem{ding2017principal}
Peng Ding and Jiannan Lu.
\newblock Principal stratification analysis using principal scores.
\newblock {\em Journal of the Royal Statistical Society: Series B (Statistical
  Methodology)}, 79(3):757--777, 2017.

\bibitem{valente2020causal}
Matthew~J Valente, Judith~JM Rijnhart, Heather~L Smyth, Felix~B Muniz, and
  David~P MacKinnon.
\newblock Causal mediation programs in r, m plus, sas, spss, and stata.
\newblock {\em Structural Equation Modeling: A Multidisciplinary Journal},
  27(6):975--984, 2020.

\end{thebibliography}
\bibliographystyle{unsrt}

\clearpage

\clearpage
\appendix

\section{More details on Separable Effects}
\label{app: more formally on separable effects}
The separable effects are defined with respect to modified treatment components, $A_Y$ and $A_S$, which are linked to the original treatment $A$ in the following way: when $A_Y$ and $A_S$ are set to the same value $a$, that is $A_Y = A_S = a$, then the counterfactual values of the event of interest $Y^{a_Y=a,a_S=a}$ and the intercurrent event $S^{a_Y=a,a_S=a}$ are equal to the counterfactual outcomes when the original treatment $A=a$, that is, $Y^{a_Y=a,a_S=a} =Y^{a} $ and $S^{a_Y=a,a_S=a} =S^{a} $. This modified treatment assumption holds when the original treatment $A$ can be decomposed into two components $A_Y$ and $A_S$. However, this assumption can also hold even if $A$ cannot be physically decomposed \cite{stensrud2020conditional}.  When the  components $A_Y$ and $A_S$ are given individually, they exclusively target certain causal pathways. For example, let $A$ be statin therapy and $Y$ be cardiovascular risk. The combination  $A_Y=1$ and $A_S=0$ can be conceived as a modified therapy (such as PCSK9 inhibitors) that, like statins, have a cholesterol lowering component $A_Y = 1$, but lacks effects on the intercurrent event such as muscle symptoms, $A_S  = 0$. To study separable effects, the pathways by which $A_Y$ and $A_S$ exert effects must be isolated from each other; in particular, we require that either $A_Y$ component exerts its effects in $Y$ or the $A_S$ component exerts effects its on $S$.  

The pathways by which $A_Y$ and $A_S$ exert effects can be described by isolation conditions \cite{stensrud2019separable, stensrud2020generalized, stensrud2020conditional}, which can be expressed as follows: 
\begin{align}
    & \text{There are no causal paths from } A_Y \text{ to } S.   \label{def: Ay partial iso} \\ 
    & \text{There are no causal paths from } A_S \text{ to } Y.   \label{def: Ad partial iso} 
\end{align}
Condition \eqref{def: Ay partial iso} and \eqref{def: Ad partial iso} are called $A_Y$ partial isolation and $A_S$ partial isolation, respectively. These assumptions can be generalised to time-varying settings \cite{stensrud2020generalized, stensrud2020conditional, robins2020interventionist}; like Bornkamp et al. \cite{bornkamp2020principal}, here we consider (simplified) intercurrent event setting, where there is a single outcome $Y$ (possibly a survival outcome) and a single intercurrent event $S$.  

To define separable effects, we consider (hypothetical) settings where the components $A_Y$ and $A_S$ are assigned separately, and thus can be given different values. More precisely, we define the separable effect of the $A_Y$ component as
\begin{equation}
\Pr (Y^{a_Y=1,a_S}=1)\text{ vs. }\Pr
(Y^{a_Y=0,a_S}=1), \quad a_S \in \{0,1\},
\label{eq: A_Y sep app}
\end{equation}
which quantifies the causal effect of the $A_Y$ component on the risk of the outcome of under an intervention that assigns $A_S=a_S$.  Similarly,
\begin{equation}
\Pr (Y^{a_Y,a_S=1}=1)\text{ vs. }\Pr
(Y^{a_Y,a_S=0}=1), \quad a_Y \in \{0,1\},
\label{eq: A_S sep app}
\end{equation}%
quantifies the causal effect of the $A_S$ component on the event of interest under an intervention that assigns $A_Y=a_Y$. 

Under \eqref{def: Ay partial iso}-\eqref{def: Ad partial iso} the separable effect \eqref{eq: A_Y sep app} quantifies the direct effect of the treatment on the outcome of interest, but this direct effect is different from the principal stratum estimand because it is defined in the entire study population. Similarly, under \eqref{def: Ay partial iso}-\eqref{def: Ad partial iso} the separable effect \eqref{eq: A_S sep app} quantifies the indirect effect of the treatment on the outcome of interest through the intercurrent event. 

Like conventional mediation estimands \cite{robins2010alternative, robins2020interventionist}, such as natural direct and indirect effects \cite{robins1992identifiability}, the separable effect quantify the mechanisms by which the treatment affects the outcome of interest \cite{robins2010alternative, robins2020interventionist, stensrud2019separable, didelez2018defining, stensrud2020generalized}. Unlike the conventional mediation estimands, the separable effects do not require specification of any intervention on the intermediate event. It is not clear that such interventions on the intercurrent events exist in any of Examples 1-5. Furthermore, a feature of the separable effects is that they are defined with respect to modified treatments, and therefore they can be directly relevant in a drug development setting where the interest is tailoring new treatments to include/exclude certain causal pathways. 



In settings with intercurrent events, researchers often express interest in estimands \textit{conditional} on the intercurrent events. Following Stensrud et al \cite{stensrud2020conditional}, under the assumption that $A_Y$ partial isolation \eqref{def: Ay partial iso} holds, we can also define a conditional separable effect as
\begin{align*}
&\mathbb{E} ( Y^{a_Y=1,a_S}  - Y^{a_Y=0,a_S} \mid
S^{a_S}=s ).
\end{align*}
The conditional separable effect is the average causal effect of the (modified) treatment $A_Y$ on $Y$ when all individuals are assigned the other (modified) treatment $A_S=a_S$ in those individuals who do not experience the intercurrent event under $A_S=a_S$ (regardless of their value of $A_Y$). 

 The conditional separable effect is restricted to subjects who have a certain value of the intercurrent event under modified treatment $a_S$. Under the partial isolation assumption, \eqref{eq: conditional sep eff 1} is equal to $\mathbb{E} ( Y^{a_Y=1,a_S} - Y^{a_Y=0,a_S} \mid S^{a}=s)$ which targets the same subgroup considered in the PS estimand \eqref{eq: princ strat} rather than \eqref{eq: princ strat ich}. However, it also explicitly quantifies a treatment effect that acts \textit{outside} of the intercurrent event (a \textit{direct} effect on the event of interest not mediated by the intercurrent event) in this subgroup.

\section{Further elaboration on Example \ref{ex intro: ada}}
\label{app: ex ada}
Here we provide some more details on the relation between separable effects and Bornkamp et al's principal stratum effect in Example \ref{ex intro: ada}. 

Assuming that nobody can develop ADAs without receiving treatment (a monotonicity assumption) and full isolation \cite{stensrud2020conditional}, the conditioning set defined by the separable effects will coincide with an estimand defined in the union of principal strata by Bornkamp et al. \cite{bornkamp2020principal}. Yet, the motivation for choosing the principal stratum estimand is not clear unless a causal question is explicitly stated and translated to a formal estimand: the theory of separable effects is precisely created to help with this task \cite{robins2010alternative, stensrud2019separable, stensrud2020conditional}, and the reasoning about separable effects is precisely what justifies why the estimand defined in a union of principal stratum is of substantial interest. In other words, the separable effects reasoning is required to justify the use of the principal stratum estimand.

\section{SWIGs}
\label{sec: app swigs}
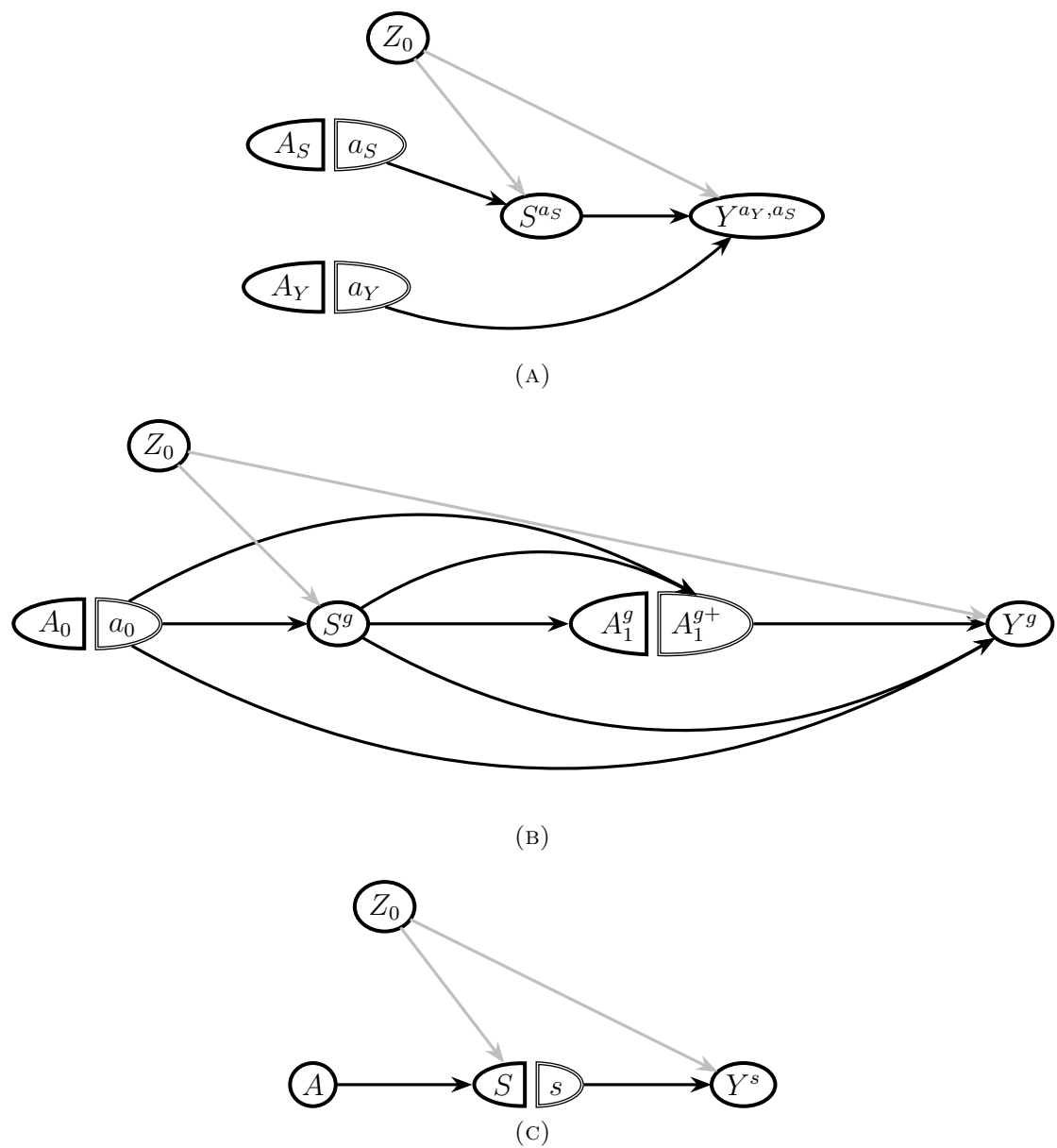
\begin{figure}
\centering
\subfloat[]{
\begin{tikzpicture}
\tikzset{line width=1.5pt, outer sep=0pt,
ell/.style={draw,fill=white, inner sep=2pt,
line width=1.5pt},
swig vsplit={gap=5pt,
inner line width right=0.5pt}};
\node[name=Ay,shape=swig vsplit] at (0,-1){
\nodepart{left}{$A_Y$}
\nodepart{right}{$a_Y$} };
\node[name=Ad, 
shape=swig vsplit] at (0,1) {
\nodepart{left}{$A_S$}
\nodepart{right}{$a_S$} };
\node[name=S,ell,  shape=ellipse] at (3,0) {$S^{a_S}$}  ;
    \node[name=Y,ell,  shape=ellipse] (Y) at (6,0) {$Y^{a_Y,a_S}$};
    \node[name=Z0,ell,  shape=ellipse] at (1,2.5) {$Z_0$};
\begin{scope}[>={Stealth[black]},
              every edge/.style={draw=black,very thick}]
	\path [->] (Ad) edge (S);
	\path [->] (Ay) edge[bend right] (Y);
    \path [->] (S) edge (Y);
    \path [->,>={Stealth[lightgray]}] (Z0) edge[lightgray]  (S);
    \path [->,>={Stealth[lightgray]}] (Z0) edge[lightgray]  (Y);
\end{scope}
\end{tikzpicture}
} \\
\subfloat[]{
\begin{tikzpicture}
\tikzset{line width=1.5pt, outer sep=0pt,
ell/.style={draw,fill=white, inner sep=2pt,
line width=1.5pt},
swig vsplit={gap=5pt,
inner line width right=0.5pt}};
\node[name=A0,shape=swig vsplit] at (-5,0){
\nodepart{left}{$A_0$}
\nodepart{right}{$a_0$} };
\node[name=A1, 
shape=swig vsplit] at (3,0) {
\nodepart{left}{$A_1^{g}$}
\nodepart{right}{$A_1^{g+}$} };
\node[name=S,ell,  shape=ellipse] at (-1.5,0) {$S^{g}$}  ;
    \node[name=Y,ell,  shape=ellipse] (Y) at (8,0) {$Y^{g}$};
    \node[name=Z0,ell,  shape=ellipse] at (-4,2.5) {$Z_0$};
\begin{scope}[>={Stealth[black]},
              every edge/.style={draw=black,very thick}]
	\path [->] (A0) edge (S);
	\path [->] (A0) edge[bend left] (3.5,0.4);
	\path [->] (A0) edge[bend right] (Y);
    \path [->] (S) edge[bend right] (Y);
    \path [->] (S) edge (A1);
    \path [->] (S) edge[bend left] (3.5,0.4);
    \path [->] (A1) edge (Y);
    \path [->,>={Stealth[lightgray]}] (Z0) edge[lightgray]  (S);
    \path [->,>={Stealth[lightgray]}] (Z0) edge[lightgray]  (Y);
\end{scope}
\end{tikzpicture}
} \\
\subfloat[]{
\begin{tikzpicture}
\tikzset{line width=1.5pt, outer sep=0pt,
ell/.style={draw,fill=white, inner sep=2pt,
line width=1.5pt},
swig vsplit={gap=5pt,
inner line width right=0.5pt}};
\node[name=S,shape=swig vsplit] at (3,0){
\nodepart{left}{$S$}
\nodepart{right}{$s$} };
\node[name=A,ell,  shape=ellipse] at (0,0) {$A$}  ;
    \node[name=Y,ell,  shape=ellipse] (Y) at (6,0) {$Y^{s}$};
    \node[name=Z0,ell,  shape=ellipse] at (1,2.5) {$Z_0$};
\begin{scope}[>={Stealth[black]},
              every edge/.style={draw=black,very thick}]
	\path [->] (A) edge (S);
    \path [->] (S) edge (Y);
    \path [->,>={Stealth[lightgray]}] (Z0) edge[lightgray]  (S);
    \path [->,>={Stealth[lightgray]}] (Z0) edge[lightgray]  (Y);
\end{scope}
\end{tikzpicture}
} \\
    \caption{ \footnotesize (A)-(C) Single World Intervention Graphs (SWIGs) with minimal labeling \cite{richardson2013single}. Superscripts denote counterfactuals. The SWIG in (A) describes a setting where conditional separable effects are identified, provided that $Z_0$ is measured. The SWIG in (B) describes a more involved setting The SWIGs are minimal labelled. The SWIG in (C) describes a setting where the dynamic treatment regime $g$ is identified conditional on $Z_0$.}
    \label{fig: swigs}
\end{figure}

It is possible to draw Single World Intervention Graphs (SWIGs) \cite{richardson2013single} to evaluate exchangeability conditions for interventionist estimands \cite{robins2020interventionist}, such as sequential trial estimands and separable effects. SWIGs are causal graphs that are related to conventional causal DAGs \cite{pearl}. Unlike a conventional DAG, the SWIG explicitly encodes a setting where the intervention of interest is instantiated (see \cite{richardson2013single}) and therefore allows us to directly evaluate counterfactual independencies for the estimand of interest. A feature of SWIGs is that these graphs only encode assumptions that are empirically falsifiable, i.e.\ testable, in a (future) study.\footnote{As identification of principal stratum estimands require additional assumptions that are empirically untestable, even in principle, we cannot use SWIGs to study exchangeability conditions for principal stratum estimands.} Splitting of nodes in a SWIG describes interventions on the nodes (i.e.\ variables) of interest, and these interventions define the counterfactual estimand. For example, the simple SWIG in Figure \ref{fig: swigs}A describes a setting where we consider counterfactual outcomes ($Y^{a_Y,a_S}$) under interventions on the treatment components $A_Y$ and $A_S$. This graph can be used to read off whether \textit{counterfactual} independencies between the treatment components $A_Y$ and $A_S$ and the counterfactual outcome $Y^{a_Y,a_S}$ hold.\footnote{Strictly speaking, this is a Single World Intervention Template \cite{richardson2013single}} The SWIGs can also be used to evaluate identification conditions for sequential trial estimands such as \eqref{eq: seq trial} in Example \ref{ex intro: ada}, as illustrated by the SWIG in Figure \ref{fig: swigs}B. Similarly, SWIGs can be used to evaluate whether the effect of $S$ in Example \ref{ex intro: survival} can be identified, which require standard conditions for identification of (conditional) causal effects \cite{hernan2020causal}.  In particular, the SWIG in Figure \ref{fig: swigs}C describes a setting where \eqref{eq: baseline cond estimand} is identified if we observe $A$ and $Z_0$.

\end{document}